\documentclass[a4]{article}

\begin{document}

% \draft command makes pacs numbers print
%\draft
%\twocolumn

\title{Yakhot's model of strong turbulence: A generalization of scaling models of turbulence. 
}
\author{Ch.  Renner, J. Peinke \\ Fachbereich Physik, Universit\"at Oldenburg} 
\date{\today} 
\maketitle

\begin{abstract}

We report on some implications of the theory of turbulence developed by V. Yakhot [V. Yakhot, Phys.  Rev.  E {\bf 57}(2) (1998)].  In particular we focus on the expression for the scaling exponents $\zeta_{n}$.  We show that Yakhot's result contains three well known scaling models as special cases, namely K41, K62 and the theory by V. L'vov and I. Procaccia [V. L'vov \& I. Procaccia, Phys.  Rev.  E {\bf 62}(6) (2000)]. The model furthermore yields a theoretical justification for the method of extended self--similarity (ESS).

\end{abstract}

% insert suggested PACS numbers in braces on next line

% \pacs{turbulence - fluid dynamics 47.27; Fokker-Planck equation- 
% stat. Physics 05.10G}

\section{Introduction}

The most challenging open problem of turbulent fluid motion is the phenomenon of small scale intermittency, i.e. the unexpected frequent occurence of very large fluctuations of the local velocity over small spatial distances. These fluctuations on a certain length scale $r$ are usually investigated by means of the longitudinal velocity increment $u(r)$, the difference of the velocities at two points in space separated by the distance $r$: $u(r) = v(x+r)-v(x)$ (here, $v$ is the component of the velocity fluctuations in direction of the separation vector ${\bf r}$). The statistics of $u(r)$ is commonly characterized by means of the moments $S^{n}(r) = \left< \, u(r)^n \, \right>$,  the so-called velocity structure functions.

Even though the first theory of fully developed turbulence dates back to 1941 \cite{K41}, a commonly accepted model for the structure functions has yet not been found (see \cite{Standardzitate} for an introduction into the field).  The only exception is the third order structure function for which an exact relation can be derived directly from the Navier--Stokes equation, Kolmogorov's famous four--fifths law. For scales $r$ much smaller than the system size $L$ and larger than the dissipation length scale $\eta$ of the flow configuration, the third order structure function is a linear function of the scale $r$ \cite{Standardzitate}:
\begin{equation}
	S^{3}(r) = - \frac{4}{5} \epsilon r. \label{FourFifths}
\end{equation}
Here, $\epsilon$ is the mean rate of energy dissipation within the flow.

In the light of the exact result (\ref{FourFifths}), Kolmogorov based his models on the hypothesis that the structure functions follow power laws in $r$
\begin{equation}
	S^{n}(r)  \propto r^{\zeta_{n}}, \label{GeneralScaling}
\end{equation}
where the $\zeta_{n}$ are the so--called scaling exponents.

Assuming a cascading process within the turbulent flow which transports energy from large to small scales with a constant rate of energy transfer, Kolmogorov \cite{K41} predicted a simple linear dependence of the scaling exponents on their argument $n$ (furtheron refered to as K41):
\begin{equation}
	\zeta_{n} =  \frac{n}{3}.  \label{K41Scaling}
\end{equation}
However, as noted by Landau \cite{Landau}, there is reason to doubt the assumption of a constant rate of energy transfer inbetween scales. Taking into account Landau's criticism, Kolmogorov \cite{K62} proposed an advanced model for the energy transport from large to small scales which, in better agreement with experimental results, predicts a quadratic dependence of the scaling exponents on $n$ (this result will furtheron be refered to as K62):
\begin{equation}
	\zeta_{n} = \frac{n}{3} - \frac{n(n-3)}{2} \delta.  \label{K62Scaling}
\end{equation}
For $n=2$ it follows from (\ref{K62Scaling}) that $\zeta_{2} = \frac{2}{3} + \delta$. The parameter $\delta$ thus describes a correction to the K41--value of $\frac{2}{3}$ and will furtheron be denoted as  anomaly parameter \cite{CommentI}. Its value is not given by the model but has to be determined experimentally, the commonly accepted value being $\delta = 0.029 \pm 0.004$ \cite{Arneodoetal}. 

This parameter has gained quite some importance within the theoretical framework recently developed by V. L'vov and I. Procaccia \cite{Lvov}. Investigating the behaviour of multiscale correlation functions for the case that two or more of the scales approach each other, these two authors derive an expression for the $\zeta_{n}$ which is of second order in $\delta$:
\begin{equation}
	\zeta_{n}  =  \frac{n}{3}  - \frac{n(n-3)}{2} \delta \left\{ 
	1+2\delta (n-2) b \right\} , \label{LvovExponents}
\end{equation}
where $b$ is of order $-1$ with a weak dependence on details of the calculations.

In the sequel, the theory propsed by V. Yakhot \cite{Yakhot} will be discussed with respect to these three models. In particular, it is shown that equations (\ref{K41Scaling}--\ref{LvovExponents}) are obtained as the zero, first and second order Taylor expansions of the scaling exponents derived by Yakhot. The theory also fixes the parameter $b$ in equation (\ref{LvovExponents}). Furthermore, it is shown that the theory provides a justification for the probably most important experimental technique to measure scaling exponents, namely the extended self--similarity (ESS) \cite{ESS}.

The paper is organized as follows. Section \ref{Repetition} summarizes some important results of Yakhot's theory while section \ref{results} gives the main results of our considerations. A discussion in section \ref{discussion} will finally conclude the paper.

\section{Some results of Yakhot's model} \label{Repetition}

Following ideas by A.M. Polyakov \cite{Polyakov}, Yakhot proposed a model of fully developed turbulence in the limit of very large Reynolds number which is based on an analytical treatement of the Navier--Stokes equation \cite{Yakhot}. For the case of turbulent flows definded by some large scale boundary condition (a typical example is free jet turbulence), he derived the following equation for the scale dependence of the probabiltiy density function $p(u)$ of the velocity increment:
\begin{equation}
	B \frac{\partial p}{\partial r} - \frac{\partial}{\partial u} 
	\left( u \frac{\partial p}{\partial r} \right) = - \frac{A}{r} 
	\frac{\partial}{\partial u} \left( u p \right) + \frac{\sigma }{L} 
	\frac{\partial^2}{\partial u^2} \left( u p \right). \label{PDFEquation}
\end{equation}
Here, $A$ and $B$ are two parameters that are not fixed by the theory and $\sigma = \sqrt{\left< v^2 \right>}$ is the rms of the velocity fluctuations $v$ (with $\left< v \right>=0$).

By multiplication of eq. (\ref{PDFEquation}) with $u^n$ and subsequent integration with respect to $u$, the equation for the structure functions $S^n(r) = \int u^n p(u) du$ can be derived. The resulting equation reads:
\begin{eqnarray}
	r \frac{\partial}{ \partial r} S^{n}(r) & = & \zeta_{n} S^{n}(r) + \sigma z_{n} \frac{r}{L} S^{n-1}(r), \label{SunEquation} \\
	\zeta_{n} & = & \frac{An}{B+n} \label{ZetaN1},  \\
	z_{n} & = & \frac{n(n-1)}{B+n}. \label{ZN1}
\end{eqnarray}
In the limit of small scales $r  \ll L$, the second term on the rhs of equation (\ref{SunEquation}) becomes neglible. In this limit, the solutions of the resulting equation are simple power laws:
\begin{equation}
	S^{n}(r) = C_{n} r^{\zeta_{n}}. \label{SmallScaleSuns}
\end{equation}
The four--fifths law (\ref{FourFifths}) imposes the condition $\zeta_{3}=1$ on the scaling exponents which, inserted into equation (\ref{ZetaN1}), leads to:
\begin{equation}
	A = \frac{B+3}{3}. \label{AvonB}
\end{equation}
The scaling exponents can thus be expressed as a function of only one yet unknown parameter:
\begin{equation}
	\zeta_{n} =  \frac{n}{3} \frac{B+3}{B+n}. \label{ZetaN}
\end{equation}
Taking into account both terms on the rhs of eq. (\ref{SunEquation}), it is furthermore possible to determine the parameter $B$. For the second order structure function the last term on the rhs of eq. (\ref{SunEquation}) vanishes on all scales (as $S^{1}(r)=0$) and the general solution is a power law in $r$: 
\begin{equation}
	S^{2}(r) = C_{2} r^{\zeta_{2}} \label{Su2I}
\end{equation}
At the integral length scale $L$ the velocity fluctuations are uncorrelated. This condition fixes the integration constant $C_{2}$ in eq. (\ref{Su2I}) and the final solution for the second order structure function is \cite{Yakhot}:
\begin{equation}
	S^{2}(r) = 2 \sigma^2 \left( \frac{r}{L} \right)^{\zeta_{2}}.  \label{Su2}
\end{equation}
Inserting this result into the equation for the third order structure function leads to:
\begin{eqnarray}
	r \frac{\partial}{\partial r}S^{3}(r) & = & \zeta_{3} S^{3}(r) + \sigma z_{3} 
	\frac{r}{L} S^{2}(r) \nonumber \\
	 & = & S^{3}(r) + 2 \sigma^3 z_{3} \left( \frac{r}{L} 
	 \right)^{\zeta_{2}+1} . 
\end{eqnarray}
The general solution for this equation is:
\begin{eqnarray}
	S^{3}(r) & = & C_{3} r + 2 \sigma^3 \frac{z_{3}}{\zeta_{2}} \left( 
	\frac{r}{L} \right)^{\zeta_{2}+1}.
\end{eqnarray}
The four--fifths law (\ref{FourFifths}) immediately determines the integration constant $C_{3}$ and the final result for the third order structure function is:
\begin{eqnarray}
	S^{3}(r) & = & - \frac{4}{5} \epsilon r + 2 \sigma^3 \frac{z_{3}}{\zeta_{2}} \left( 
	\frac{r}{L} \right)^{\zeta_{2}+1} \nonumber \\
	& = & - \frac{4}{5} \epsilon r + 18 \sigma^3 \frac{B+2}{\left( B+3 
	\right)^2} \left( 
	\frac{r}{L} \right)^{\zeta_{2}+1} . \label{Su3}
\end{eqnarray}
The parameter $B$ can be determined from the condition that odd order moments vanish at the integral length scale, i.e. 
$S^{3}(L)=0$. As furthermore $\epsilon L \approx \sigma^3$, it follows from eq. (\ref{Su3}) that
\begin{equation}
	\frac{4}{5} \approx 18 \frac{B+2}{\left( B+3 \right)^2} 
	\label{BBestimmung}
\end{equation}
which is solved by $B \approx 20$.  With this value, the scaling exponents (\ref{ZetaN}) are, within the errors, indistinguishable from the best experimental values \cite{Yakhot}.

\section{A Reformulation} \label{results}

An interesting result can be derived from Yakhot's theory by replacing the parameter $B$ in the above equations by the scaling anomaly $\delta$. In accord with equations (\ref{K62Scaling}) and (\ref{LvovExponents}) we define $\delta$ as:
\begin{eqnarray}
	\zeta_{2} & = & \frac{2}{3} + \delta. \label{DeltaDef}
\end{eqnarray}
On the other hand, we have from equation (\ref{ZetaN}) that
\begin{equation}
	\zeta_{2} = \frac{2}{3} \frac{B+3}{B+2}.  \label{Zeta2}
\end{equation}
%Comparison of equations (\ref{DeltaDef}) and (\ref{Zeta2}) finally leads to:
%\begin{equation}
%	B = 2  \frac{1-3 \delta}{3 \delta}.
%\end{equation}
%Inserting this result into Yakhot's formula (\ref{ZetaN}) for the scaling exponents yields the follwing 
%expression for the $\zeta_{n}$:
Using these two relations to replace the parameter $B$ in eq. (\ref{ZetaN}) leads to the following expression for the scaling exponents:
\begin{eqnarray}
	\zeta_{n} = \frac{n}{3} \frac{2+3 \delta}{2+3(n-2) \delta}. \label{MyZetaN}
\end{eqnarray}
As $\delta$ is a small parameter, it is possible to expand the $\zeta_{n}$ in a Taylor series around $\delta = 0$. Up to order two in $\delta$ this expansion reads:
\begin{eqnarray}
	\zeta_{n} & \approx & \frac{n}{3}  \left\{ 1- \frac{3}{2}(n-3) \delta + \frac{9}{4}(n-3)(n-2) \delta^2  \right\} \nonumber \\
	& = & \frac{n}{3} - \frac{n(n-3)}{2} \delta \left\{ 1 + 2 \delta (n-2) \left( -\frac{3}{4} \right) \right\}. \label{YakhotApprox}
\end{eqnarray}
This is exactly the expression derived by L'vov and Procaccia, equation (\ref{LvovExponents}), and in agreement with the order of magnitude they find we obtain $b = -3/4$.

It is also interesting to reformulate the equation for the third order structure function, equation (\ref{Su3}), in terms of $\delta$:
\begin{eqnarray}
	S^{3}(r) & = & - \frac{4}{5} \epsilon r + 12 \sigma^3 \delta \left( \frac{3}{2 + 3 \delta} \right)^2 \left( \frac{r}{L} 	\right)^{\zeta_{2}+1}  \nonumber \\
	& \approx &  - \frac{4}{5} \epsilon r + 27 \delta \sigma^3 \left( \frac{r}{L} \right)^{\zeta_{2}+1}. \label{Su3Delta}
\end{eqnarray}
In the second line we neglected terms of order $\delta^2$ and higher. Setting $S^{3}(L)=0$ we finally obtain 
\begin{eqnarray}
	\delta \approx \frac{4}{135} \frac{\epsilon L}{\sigma^3} \approx \frac{4}{135} \approx 0.03, \label{Delta}
\end{eqnarray}
in full agreement with the experimental value of $\delta = 0.029 \pm 0.004$ \cite{ExactSolution}. 

Yakhot's theory also allows for an interesting statement on the moments  $T^{n}(r) = \left< \left| u(r) \right|^{n} \right>$ of the absolute value of $u(r)$. From equation (\ref{PDFEquation}) for the pdf $p(u)$, the corresponding equation for the pdf $p(-u)$ of the negative increment can easily be derived. Adding the equations for $p(u)$ and $p(-u)$ and using the definitions
\begin{eqnarray}
	p^{+}(u) & := & p(u) + p(-u) \nonumber \\
	p^{-}(u) & := & p(u) - p(-u)
\end{eqnarray}
we obtain:
\begin{eqnarray}
	B \frac{\partial p^{+}}{\partial r} - \frac{\partial}{\partial u} 
	\left( u \frac{\partial p^{+}}{\partial r} \right) = & - & \frac{A}{r} 
	\frac{\partial}{\partial u} \left( u p^{+} \right)  \nonumber \\
	& + & \frac{ \sigma }{L} 
	\frac{\partial^2}{\partial u^2} \left( u p^{-} \right). \label{pPlusEq}
\end{eqnarray}
As the $T^{n}(r)$ can be expressed by means of $p^{+}$:
\begin{eqnarray}
	T^{n}(r) %& = & \int\limits_{0}^{+ \infty} u^n \left( p(u) + p(-u) \right) du \nonumber \\
	& = & \int\limits_{0}^{+ \infty} u^n  p^{+}(u)  du  ,
\end{eqnarray}
the equation for these moments can be obtained by multiplying equation (\ref{pPlusEq}) with $u^n$ and subsequently integrating with respect to $u$. The result is:
\begin{eqnarray}
	r \frac{\partial}{ \partial r} T^{n}(r) = \zeta_{n} T^{n}(r) + \sigma z_{n} \frac{r}{L} t^{n-1}(r), \label{TunEquation}
\end{eqnarray}
where we defined the moments $ t^{n}(r)$ as $ t^{n}(r)  =\int_{0}^{+\infty} u^n p^{-}(u)  du $ (these quantities can be thought of as generalized odd order moments).
	
Equations (\ref{SunEquation}) and (\ref{TunEquation}) for the odd order moments $T^{2n+1}(r)$ and $S^{2n+1}(r)$ differ in the last term on the rhs. While the equation for $S^{2n+1}(r)$ incorporates the large even order moment $S^{2n}(r)$, the moment $t^{2n}(r)$ contained in the equation for $T^{2n+1}(r)$ is considerably smaller and even tends to zero as $r$ approaches the integral length scale. In other words: While this term dominates the behaviour of the $S^{2n+1}(r)$ for large scales, it has a much weaker influence on the moments $T^{2n+1}(r)$. According to this result, odd order moments of absolute velocity increments can in a much better approximation and over a wider range of scales be described as power laws in $r$. This finding is in full agreement with experimetnal results \cite{ESS}. Equation (\ref{TunEquation}) furthermore tells us that the scaling exponents determined from the moments $T^{2n+1}(r)$ are indeed identical to the exponents that determine the scaling of the structure functions $S^{2n+1}(r)$ in the limit of smale scales.
Things even improve if, as it is the case for the extended self--similarity, the relative scaling of structure functions is considered. Basically, this method assumes a linear dependence between the logarithms of structure functions of various order \cite{ESS} where the slope is given by the scaling exponents. The ESS--property can be formulated as:
\begin{equation}
	\frac{\partial \ln T^{n}(r) }{\partial \ln r} = \frac{\zeta_{n}}{\zeta_{p}} \frac{\partial \ln T^{p}(r)}{\partial \ln r}. \label{ESSDef}
\end{equation}
An equation quite similar to the ESS--property (\ref{ESSDef}) can be derived from Yakhot's model. Equation (\ref{TunEquation}) for the $T^{n}(r)$ can be rewritten as
\begin{eqnarray}
	1 & = & \frac{1}{\zeta_{n}} \left\{ \frac{\partial \ln T^{n}(r) }{\partial \ln r} - \sigma z_{n} \frac{r}{L} \frac{t^{n-1}(r)}{T^{n}(r)} 	\right\} \nonumber \\
	& = & \frac{1}{\zeta_{p}} \left\{ \frac{\partial \ln T^{p}(r) }{\partial \ln r} - \sigma z_{p} \frac{r}{L} \frac{t^{p-1}(r)}{T^{p}(r)} 	\right\}
\end{eqnarray}
which finally yields:
\begin{equation}
	\frac{\partial \ln T^{n}(r) }{\partial \ln r} = \frac{\zeta_{n}}{\zeta_{p}} \frac{\partial \ln T^{p}(r)}{\partial \ln r} +  \sigma z_{n} 	\frac{r}{L} \left\{ \frac{t^{n-1}(r)}{T^{n}(r)} - \frac{p-1}{n-1} \frac{t^{p-1}(r)}{T^{p}(r)} \right\}. \label{ESSEq}
\end{equation}
Apart from the last term on the right hand side, this equation is just the ESS--property as expressed by equation (\ref{ESSDef}). It is furthermore found that this last term not only inlcudes the small quantities $t^{n-1}(r)/T^{n}(r)$ but even the difference of two of these small terms. It thus turns out that the extended self--similarity minimizes the influence of this term and in fact measures the scaling exponents that describe the small--scale structure of turbulence.

%Another interesting consequence of equation (\ref{ESSEq}) is that the influence of this last term is minimal if %$n$ and $p$ do not differ too much. Even better results for the scaling exponents might therefore be %achievable if, instead of expressing all the $\zeta_{n}$ as functions of $\zeta_{3}$, higher order exponents %are sucessivley determined via a series of relative exponents with $n-p=1$, i.e. the relative exponents  %$\zeta_4/\zeta_3$, $\zeta_5/\zeta_4$, $\zeta_6/\zeta_5$,  etc.

\section{Conclusion and Comments} \label{discussion}

From a theoretical point of view the most interesting of our results is the finding that Yakhot's model unifies three of the most important theories of strong turbulence: It not nonly contains K41 and K62 as limiting cases but also the scaling exponents predicted by L'vov and Procaccia. Besides, Yakhot's theory applies to a much wider range of scales and physical phenomenona: The results presented here were obtained  by taking two limits, the limit of small scales $r$ and the limit of a small anomaly parameter $\delta$ (see \cite{Yakhot} for a discussion of the extreme cases $B=0$ and  $B \rightarrow +\infty$).

%It is furthermore worth to note that this result is obtained by taking two limits, the limit of small scales $r$ and the limit of a small anomaly parameter $\delta$; see \cite{Yakhot} for a discussion of the extreme cases $B=0$ (which corresponds to $\delta=\frac{1}{3}$) and  $B \rightarrow +\infty$ (or $\delta=0$).

%Yakhot's model thus not only unifies three of the most important theories of turbulence but also applies to a wider range of scales and physical phenomena; see \cite{Yakhot} for a discussion of the limiting cases $B=0$ and $B \rightarrow +\infty$ (or $\delta=\frac{1}{3}$ and $\delta=0$, respectively).

Another important result follows from Yakhot's model for the extended self--similarity and the use of moments of absolute velocity increments.The theory gives evidence that both methods in fact allow for a precise and reliable determination of the scaling exponents that describe the small-scale behaviour of the structure functions.

Finally, we would like to point out that the model equations such as eq. (\ref{Delta}) predict a dependence of the scaling parameter $\delta$ on the large scale properties of the flow. This dependence, though obviously being weak, %as the relation $\epsilon L \approx \sigma^3$ can be expected to hold in good approximation, 
is in contradiction to the commonly assumed universal properties of small scale turbulence. lt is, however, in agreement with recent experimental findings \cite{ZetaZweiVonRe, PRL}.

\subsection*{Acknowledgments} 
We gratefully acknowledge fruitful discussions with R. Friedrich and M. Siefert.


\begin{thebibliography}{}
\bibitem{CommentI} It should be noted that equation (\ref{K62Scaling}) is usually not formulated using the 	second 	order exponent's anomaly $\delta$ but rather the sixth's order anomaly $\mu$ (where $\mu = 9 	\delta$).
\bibitem{ExactSolution} The exact solution including all orders of $\delta$ in eq. (\ref{Su3Delta}) is:
	\begin{eqnarray}
		\delta_{\pm} = \frac{2}{3 \kappa} \left\{ 1 \pm \sqrt{1-\kappa^2} \right\}, \nonumber
	\end{eqnarray}
	where $\kappa = \frac{4}{45c-4}$ and $c=\frac{\sigma^3}{\epsilon L}$. Setting $c=1$ we obtain $\delta_{-	}=0.0326$ 	and expanding the square root yields $\delta_{-} \approx \frac{\kappa}{3} \approx 	\frac{4}{123} = 0.0325$. The latter 	result, $\delta \approx \frac{ \kappa}{3}$, might serve as a starting 	point for experimental investigations concerning the dependence of $\delta$ on $c$.
\end{thebibliography}
\end{document}